# Voltage-Controlled Skyrmionic Interconnect with Multiple Magnetic Information Carriers

Runze Chen[1, *], Yu Li[1], Vasilis F. Pavlidis[2], and Christoforos Moutafis[1, &]

[1] *Nano Engineering and Spintronic Technologies (NEST) research group, Department of Computer Science, The University of Manchester, Manchester M13 9PL, United Kingdom*
[2] *Advanced Processor Technologies (APT) research group, Department of Computer Science, The University of Manchester, Manchester M13 9PL, United Kingdom*

[*]runzechen.spin@outlook.com
[&]christoforos.moutafis@manchester.ac.uk

## Abstract

Magnetic skyrmions have been in the spotlight since their observation in technologically relevant systems at room temperature. More recently, there has been increasing interest in additional quasiparticles that may exist as stable/metastable spin textures in magnets, such as the skyrmionium and the anti-skyrmionite (i.e., a double-antiskyrmion-skyrmion pair) that have distinct topological characteristics. The next challenge and opportunity, at the same time, is to investigate the use of multiple magnetic quasiparticles as information carriers in a single device for next generation nanocomputing. In this paper, we propose a spintronic interconnect device where multiple sequences of information signals are encoded and transmitted simultaneously by skyrmions, skyrmioniums, and anti-skyrmionites. The proposed spintronic interconnect device can be pipelined via voltage-controlled-magnetic-anisotropy (VCMA) gated synchronisers that behave as intermediate registers. We demonstrate theoretically that the interconnect throughput and transmission energy can be effectively tuned by the VCMA gate voltage and appropriate electric current pulses. By carefully adjusting the device structure characteristics, our spintronic interconnect device exhibits comparable energy efficiency with copper interconnects in mainstream CMOS technologies. This study provides fresh insight into the possibilities of skyrmionic devices in future spintronic applications.

# 1. Introduction

Magnetic skyrmions are non-trivial particle-like spin textures stabilised in non-centrosymmetric bulk magnets and thin magnetic films with broken inversion symmetry. This behaviour is made possible by the interactions favouring nonparallel magnetization orientation, such as Dzyaloshinskii-Moriya interaction (DMI) and dipolar coupling [1–3]. Skyrmionic spin textures can be characterized by topological indices, such as the skyrmion number, which counts how many times the vector field configuration wraps around a unit sphere and reflects the topological charge they may be endowed with. The skyrmion number is defined as:

$$N_{\text{sk}} = \frac{1}{4\pi} \int \boldsymbol{m} \cdot \left( \frac{\partial \boldsymbol{m}}{\partial x} \times \frac{\partial \boldsymbol{m}}{\partial y} \right) dxdy, \tag{1}$$

where $N_{\text{sk}} = \pm 1$ accounts for the case of magnetic skyrmions, and the sign reflects their polarity. The skyrmionium, a skyrmion-antiskyrmion pair, is a magnetic quasiparticle with vanishing topology $N_{\text{sk}} = 0$, which has been proposed as advantageous for racetrack memory applications [4]. More recently, several studies suggest that nanomagnets can host a plethora of topological quasiparticles, including both theoretical calculations [5] and experimental demonstrations such as skyrmion bags in liquid crystals [6] and skyrmion bundles in chiral magnets [7].

Based on these results, the skyrmionic quasiparticles can be promising candidates for future low-power and low-temperature computing due to their non-volatility, nanoscale size, and ease of manipulation [8]. To date, the use of skyrmionic quasiparticles has been proposed in conventional computing and emerging computational paradigms, such as skyrmionic transistors [9], skyrmionic logic gates [10–12], skyrmionic racetrack memory [13–15], spintronic nano-oscillators [16], skyrmionic resonant diodes [17], skyrmion neuromorphic computing [18,19], and reservoir computing [20], where Ref. [19] demonstrates experimental results, and the others are simulation results. However, apart from information processing, the efficient information transmission with spintronic devices, e.g., spintronic interconnect devices and multiplexers, has not received the required attention. Considering that interconnect energy and latency often is the bottleneck of modern computing systems, novel and effective interconnect is an indispensable requirement for the adoption of any emerging technology.

In complementary metal-oxide-semiconductor (CMOS) integrated circuits, interconnects link two or more circuit elements (i.e., transistors) electrically [21], where copper wires are commonly utilised. However, the power spent on copper interconnects can exceed the energy spent for computation, which becomes a critical challenge towards delivering exascale performance at a reasonable power budget [21]. It is therefore desirable to address the problem by exploiting emerging technologies. Optical interconnects are promising energy-efficient solutions for long-distance and parallel data transmission [22]. Since the last decade, silicon photonic interconnects with the ability to use CMOS-compatible fabrication have made significant strides that can benefit many applications, e.g., data centres, high-performance computing, and sensing [23]. Here, we explore alternative non-volatile and energy efficient information transfer, through signal multiplexing using spintronic devices. In our recent work, a prototype of notch-based interconnect device was proposed [24], which can perform topological filtering that enables signal multiplexing utilising sequences of magnetic skyrmions and skyrmioniums. The notch-based nanotrack design has been frequently utilised in recent numerical studies [10,15,24]. However, to avoid insidious risks that include likelihood of pinning and annihilation at notches/edges and to achieve effective performance tunability require the design of more realistic and robust spintronic interconnect devices.

We therefore propose a spintronic interconnect device in a ferromagnetic nanotrack with voltage-controlled magnetic anisotropy (VCMA) gates, and we deliver a systematic study of its operation and performance via theoretical calculations and micromagnetic simulations. The proposed interconnect device utilises multiple magnetic quasiparticles as information carriers, i.e., the topologically non-trivial skyrmions and anti-skyrmionites (a double-antiskyrmion-skyrmion pair) and the topologically trivial skyrmioniums. This approach exhibits superior thermodynamic stability than notch-based devices, as evidenced by our thermal stability analyses on the nanotrack. We then demonstrate the effective tunability of the device performance and information transmission energy. Finally, the proposed pipelined spintronic interconnect, achieved via VCMA gates, shows a comparable energy efficiency with copper interconnects in CMOS by carefully optimising the nanotrack geometry.

## 2. Results

### 2.1 VCMA-based spintronic interconnect device

The proposed spintronic interconnect device is schematically depicted in Fig. 1. The spintronic interconnect consists of four key modules: 1) The three write heads in the left of Fig. 1(a), which nucleate the corresponding magnetic quasiparticles according to the signal from the interface circuits, as a three-branch encoder; 2) The ferromagnetic/heavy metal heterostructure nanotrack supporting the propagation of the carrier streams via the spin-orbit torque (SOT); 3) VCMA-based gates distributed evenly on the nanotrack serving as synchronisers; 4) The three read heads detecting the presence/absence of a quasiparticle in the corresponding branch.

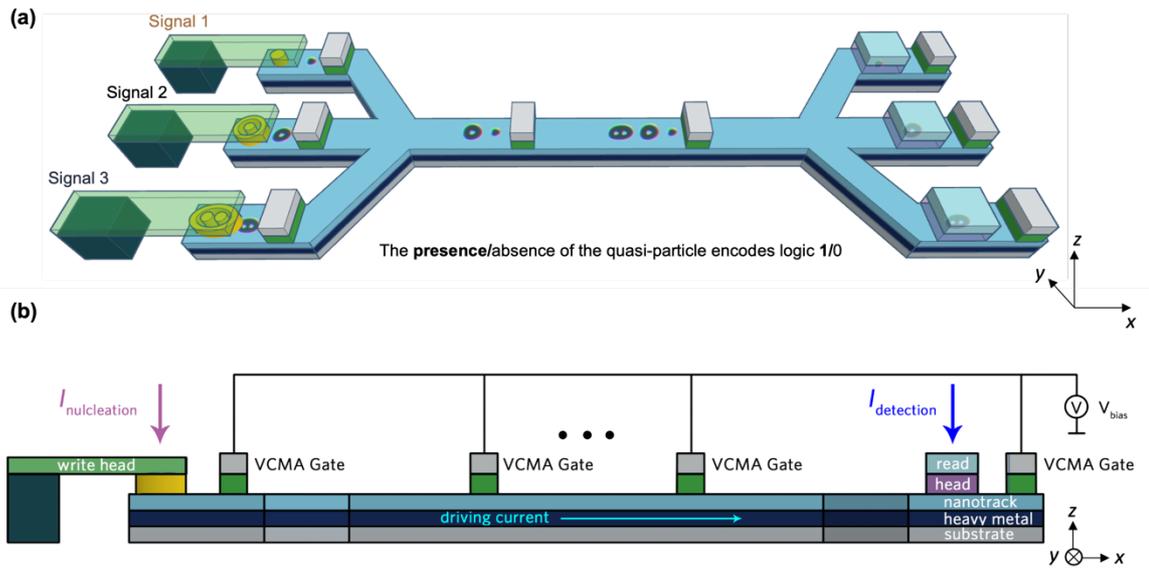

**Figure 1. Schematic drawing of the proposed voltage-controlled spintronic interconnect device:** (a) perspective view of the device; (b) cross-sectional view. The proposed device comprises spin-transfer torque (STT) write heads, VCMA-controlled gates, nanotrack, and MTJ-based read heads.

As shown in Fig. 1, there is a three-branch multiplexer on the left side and a three-branch demultiplexer on the right side. Signals are encoded (nucleated) through the three-branch multiplexer into the nanotrack and decoded (detected) via the three-branch demultiplexer. The SOT achieves the transmission, and the intrinsic skyrmion Hall effect (SkHE) enables an automatic topological filtering process. The device can work bidirectionally due to the symmetry of this design and the propagation properties of skyrmionic quasiparticles; our interconnect device may also be relevant to reversible computing [25]. Because the quasiparticles present slightly different velocities under SOT on the nanotrack [26], it is crucial to introduce synchronisers in the device both for precise detection and information integrity [24]. It should be noted that the magnetic tunnelling junction (MTJ) reading heads are illustrated for signal detection in Fig. 1 merely as an example. Besides MTJ detectors [27], we could also utilise other detection techniques, such as measuring the topological Hall effect (THE) [4,28] and hinge spin polarization [29]. Note that the hinge spin polarization was first proposed in magnetic topological insulators, but the feature is general and, therefore, relevant to the ferromagnetic and the anti-ferromagnetic phases as a potential detection method.

The perpendicular magnetic anisotropy (PMA), also called perpendicular uniaxial anisotropy, can be locally modulated by applying a voltage in thin films [9,14,30–35], which is known as the voltage-controlled magnetic anisotropy (VCMA) effect and provides great potential in practical applications. The VCMA effect was first reported in 3d-transition ferromagnetic materials in 2007 [35], where a coercivity change was observed in 2–4 nm thick FePt and FePd films immersed in a liquid electrolyte. Opposite trends in the change in coercivity depending on the applied voltage were also observed. Surprisingly, T. Maruyama et al. reported that an electric field of about 100 mV nm$^{-1}$ can change PMA of a bcc Fe(001)|MgO(001) junction by 40% with the VCMA efficiency $\vartheta$ of 210 fJ V$^{-1}$m$^{-1}$ at room temperature [34]. The VCMA efficiency $\vartheta$ is defined as the ratio between the voltage-induced change of total uniaxial anisotropy near the interface (in units of J m$^{-2}$) and the applied electric field (in units of V m$^{-1}$) [36]. The micromagnetic simulation of the VCMA effect in this work is based on a linear relationship describing the effective contribution of the applied electric field to the uniaxial magnetocrystalline anisotropy constant [14,32–34]:

$$K_{\mathrm{u1v}} = K_{\mathrm{u1}} + \frac{\vartheta E_{\mathrm{b}}}{t}, \qquad (2)$$

where $E_{\mathrm{b}}$ is the applied electric field on the VCMA gate (in units of V m$^{-1}$), $K_{\mathrm{u1v}}$ is the resulting anisotropy constant after the electric field application, $K_{\mathrm{u1}}$ is the initial uniaxial anisotropy constant (in units of J m$^{-3}$), and $t$ is the thickness of the FM layer. The region with higher PMA can provide an extra barrier, while the one with lower PMA offers a potential well. In the remaining of this paper, we refer to the region with higher/lower PMA as a VCMA barrier/well. It should be noted that the material underneath VCMA gates can be pre-set with higher PMA values, which can be achieved by locally modulating material properties during the deposition process [12]. As a result, the quasiparticles will be stopped by the barrier without voltage supply and pass through the voltage-gated region with reduced PMA by applying negative bias voltage $V_{\mathrm{b}}$. Therefore, no voltage supply of VCMA gates is required between clock cycles in this design, significantly reducing the amount of leaked charge and static power dissipation.

In this work, magnetic quasiparticles are simulated in a single FM|HM bilayer heterostructure without thermal effects. We have already demonstrated that, by introducing a tailored magnetic multilayer structure interconnect, skyrmionic quasiparticles under finite temperature can exhibit stable behaviour similar to that in thermal-free systems [18,24]. Therefore, simulations on the thermal fluctuation at finite temperature are not considered in the main results of this paper. Micromagnetic simulations are performed using the open-source package MuMax$^3$ [37]. Calculations of equilibrium states of topological spin textures and minimum energy paths (MEPs) are performed using Fidimag [38], where the nudged elastic band method (NEBM) method is used for the calculation of MEPs between the equilibrium states. With this method, the energy barriers within transitions can be quantified. Parameters and the detailed configurations utilised in the simulations are introduced in Methods.

## 2.2 Topological properties of the magnetic quasiparticles

When a skyrmion moves along the nanotrack driven by the electric current, the Magnus force stemming from the spin precession causes its movement along a trajectory at an angle to the direction of the applied current, which is the well-known skyrmion Hall effect (SkHE) [39,40]. As for other magnetic quasiparticles, a similar deflection angle related to their topological charge can be obtained [7,26]. The SkHE is usually deemed harmful to skyrmionic devices because it affects the device performance and robustness, leading to the annihilation of skyrmions at the boundaries [13]. Several strategies are thus proposed to suppress the SkHE, such as stabilising skyrmions in synthetic antiferromagnets (SAF) structures [41], adding high-$K$ materials at boundaries [10,17], moderating racetrack structure [42], moderating the spin Hall angle [43]. However, in this work, we take a different approach to exploit the SkHE, rather than circumvent it, to enable device implementation with multiple information carriers. The proposed spintronic interconnect can conduct automatic demultiplexing at the decoder (three-branch structure shown on the right side of Fig. 1(a)), which exploits topological filtering arising inherently from the SkHE. In this way we realise the combination of salient topological properties of magnetic quasiparticles with their practical applications.

We performed micromagnetic simulations of a skyrmion, skyrmionium, and anti-skyrmionite in a nanotrack under SOT. Electric current is applied in the HM layer underneath the FM layer such that a spin current will be injected perpendicular to the FM plane with the spin polarisation in $+y$ direction due to the spin Hall effect. Note that the skyrmion in our simulations carries a topological charge $N_{sk} = -1$ reflected by the negative polarity of the skyrmion core (spin ↓), while the anti-skyrmionite delivers a total net topological charge of $N_{sk} = +1$, which acts as an effective antiskyrmion in the system. As shown in Fig. 2(a), the skyrmion ($N_{sk} = -1$) propagates first toward $+y$ direction, and then along the direction of the driving current ($+x$), the anti-skyrmionite ($N_{sk} = +1$) propagates first toward $-y$ direction and then along the direction of the driving current ($+x$). In contrast, the topologically trivial skyrmionium propagates strictly along the $+x$ direction without any deflection. The skyrmion Hall angle $\theta_{SkHE}$ is defined as the angle between the trajectory of quasiparticles and the direction of applied current ($+x$), with positive sign when it is anticlockwise. From the simulation results, the skyrmion exhibits a larger absolute $\theta_{SkHE}$ than the anti-skyrmionite, even though they share the same absolute topological charge. To highlight the reason for this difference, we then perform a theoretical analysis on the movement of quasiparticles under electrical current.

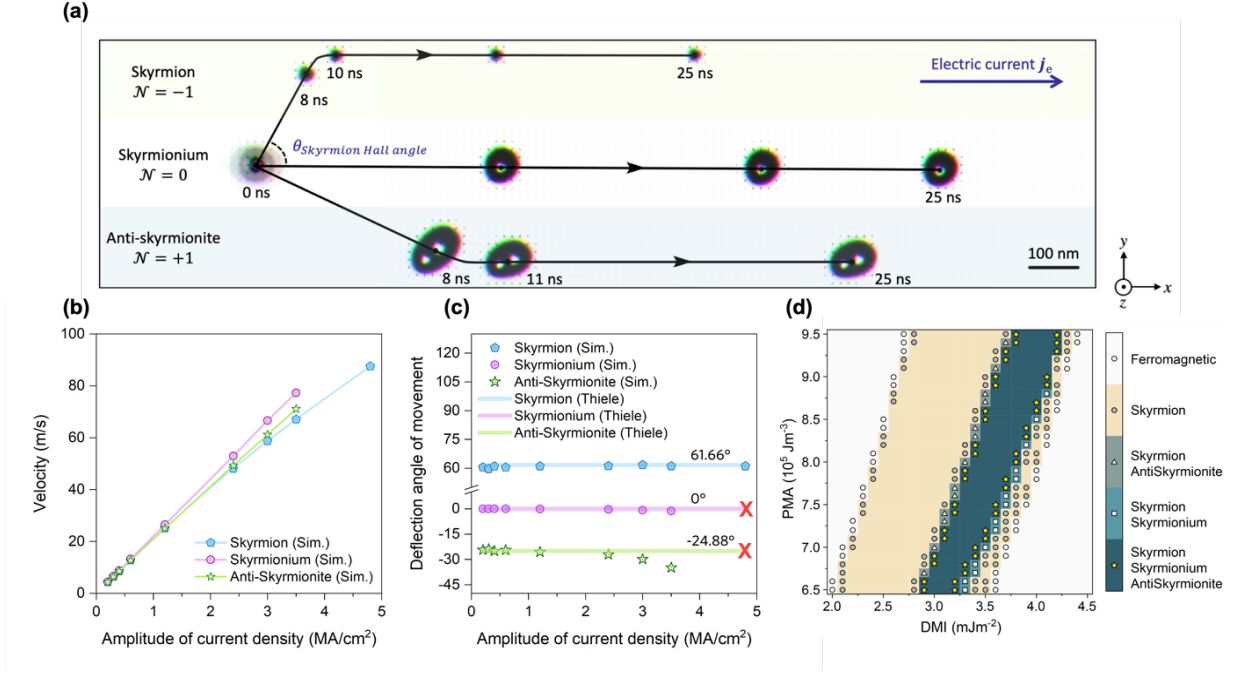

**Figure 2. Skyrmionic quasiparticles Hall effect.** (a) Micromagnetic simulation results of the propagation of a Néel skyrmion ($N_{sk} = -1$), a skyrmionium ($N_{sk} = 0$), and an anti-skyrmionite ($N_{sk} = +1$) driven by a spin current perpendicularly injected to the $x$-$y$ plane. (b) Velocities of quasiparticles propagating along the edge of the nanotrack ($x$-direction) as a function of the current density derived by simulations. (c) Hall angle of movement of the quasiparticles with respect to current densities both for simulations and theoretical calculations. The red crosses indicate that the skyrmionium and the antiskyrmionite annihilated when current densities $> 3.5 \text{ MA} \cdot \text{cm}^{-2}$. (d) Magnetic stability phase diagram of the magnetic quasiparticles. The skyrmion, skyrmionium, and anti-skyrmionite can co-exist in magnetic systems with parameters coloured in navy (the darkest green in the figure) and marked with yellow stars.

Micromagnetic simulations are performed by solving the LLG equation (see Methods), a time-dependent partial differential equation (PDE) consisting of several terms. To better comprehend the results of spin dynamics simulations, it is desirable to find an algebraic, practical description for the motion of non-collinear spin textures. By considering a magnetic quasiparticle as a rigid particle whose shape does not change significantly during the movement, the translational motion driven by the SOT can be described by a modified Thiele equation [44,45], which is an analytically solvable system of algebraic equations describing the velocity of magnetic quasiparticles,

$$\boldsymbol{G} \times \boldsymbol{v} - \alpha \boldsymbol{\mathcal{D}} \cdot \boldsymbol{v} - \mathcal{T}_{\text{SOT}} \boldsymbol{\mathcal{J}} \cdot \mathbf{m}_{\text{p}} - \nabla U(\boldsymbol{r}) = 0, \quad (3)$$

where $\boldsymbol{G} = (0, 0, -4\pi N_{sk})$ is the Gyroscopic vector with the topological charge $N_{sk}$ defined in Eq. 1. $\boldsymbol{v} = (v_x, v_y)$ is the drifting velocity of skyrmionic quasiparticle within the $x$-$y$ plane.

The first term $\boldsymbol{G} \times \boldsymbol{v}$ in Eq. 3 is the Magnus force that results in the transverse motion of skyrmions related to the spin precession, which directly results in SkHE [39]. $\alpha$ is the dimensionless Gilbert damping parameter. $\boldsymbol{\mathcal{D}}$ is the dissipative tensor and can be calculated by $\mathcal{D}_{ij} = \frac{1}{M_s^2} \iint \frac{\partial \boldsymbol{M}}{\partial x_i} \cdot \frac{\partial \boldsymbol{M}}{\partial x_j} dxdy = \begin{bmatrix} \mathcal{D}_{xx} & \mathcal{D}_{xy} \\ \mathcal{D}_{yx} & \mathcal{D}_{yy} \end{bmatrix}$, where $M_s$ is saturation magnetisation. The term $\mathcal{T}_{SOT} \boldsymbol{\mathcal{J}} \cdot \boldsymbol{m}_p$ quantifies the effect of the SOT driving the magnetic quasiparticle. $\mathcal{T}_{SOT} = \frac{\gamma_e \hbar}{2eM_s t} J_e \theta_{SH}$ is the amplitude of SOT over the quasiparticle, where $\gamma_e = \frac{\gamma_{LL}}{\mu_0} = 1.76 \times 10^{11} T^{-1} s^{-1}$ is the gyromagnetic ratio of an electron, $\hbar$ is the reduced Planck constant, $J_e$ is the electron current density, $\theta_{SH}$ is the spin Hall ratio (the ratio between the spin current and the electron current), $e$ is the electron charge, and $t$ is the thickness of the FM layer. $\boldsymbol{\mathcal{J}}$ is the driving torque tensor and can be calculated by $\mathcal{J}_{ij} = \frac{1}{M_s^2} \iint \left( \frac{\partial \boldsymbol{M}}{\partial x_i} \times \boldsymbol{M} \right)_j dxdy = \begin{bmatrix} \mathcal{J}_{xx} & \mathcal{J}_{xy} \\ \mathcal{J}_{yx} & \mathcal{J}_{yy} \end{bmatrix}$. $\boldsymbol{m}_p$ is the polarisation orientation of the spin current due to the spin Hall effect. The fourth term $\nabla U(\boldsymbol{r})$ in Eq. 3 is an extrinsic force accounting for the interaction of a magnetic quasiparticle with other non-collinearities or the nanotrack edge.

As for the initial part of the motion after applying the SOT, the quasiparticle is far from the nanotrack edge, therefore, the $\nabla U(\boldsymbol{r})$ term in Eq. 3 can be neglected. By solving Eq. 3 (details see Methods), we can obtain the skyrmion Hall angle of the quasiparticle as,

$$\tan \theta_{SkHE} = \frac{v_y}{v_x} = -\frac{4\pi N_{sk}}{\alpha \mathcal{D}_{xx}}. \tag{4}$$

After the transverse movement of the quasiparticle, supposing a significantly small current density, the quasiparticle moves along the direction of the driving current along the edge of the nanotrack, i.e., $v_y = 0$ and $v_x \neq 0$, as demonstrated in Fig. 2(a). In this condition, the $\nabla U(\boldsymbol{r})$ term in Eq. 3 will have a finite value. By solving Eq. 3 according to these constraints (details see Methods), we can extract the velocity $v_x$ as,

$$v_x = \frac{\mathcal{T}_{SOT} \mathcal{J}_{xy}}{\alpha \mathcal{D}_{xx}} \propto \frac{\mathcal{J}_{xy} J_e}{\mathcal{D}_{xx}}. \tag{5}$$

To verify the theoretical predictions from Eq. 4 and Eq. 5 derived from the Thiele equation, we performed micromagnetic simulations of an individual skyrmion, skyrmionium, and anti-skyrmionite, respectively, to obtain their velocity $v_x$ and skyrmion Hall angle $\theta_{SkHE}$ under increasing amplitude of current densities.

As shown in Fig. 2(b), the velocity of the magnetic quasiparticles changes linearly with the applied current density. The final velocity $v_x$ of the skyrmionium is larger than that of skyrmion and anti-skyrmionite. This behaviour can be verified by calculating the $\frac{\mathcal{J}_{xy} J_e}{\mathcal{D}_{xx}}$ term for each quasiparticle from Eq. 5. The skyrmionium has the most significant value of $\frac{\mathcal{J}_{xy}J_e}{\mathcal{D}_{xx}}$, resulting in the largest velocity $v_x$ along the direction of the driving current. Fig. 2(c) shows the Hall angle of movement for three magnetic quasiparticles with respect to the current densities. Discrete data points represent micromagnetic simulation results, while the solid lines are results calculated from the Thiele equation. At smaller amplitude of current densities ($< 2 \, \text{MA} \cdot \text{cm}^{-2}$), micromagnetic simulation results of three magnetic quasiparticles fit well with the results calculated from the Thiele equation. However, at higher amplitude current densities ($> 2.5 \, \text{MA} \cdot \text{cm}^{-2}$), micromagnetic simulation results of the skyrmionium and anti-skyrmionite start to derail from the theoretical predictions, while the skyrmion continues to follow the theoretical calculation of Thiele equation. This divergence can be explained by the shape distortion and rotation of the skyrmionium and anti-skyrmionite under high current densities [7,46], which would be expected for the rigid particle approximation to show its limits for larger quasi-particles. As verification, we checked each quasiparticle's topological charge and dissipation tensor with increasing current densities. The topological charge roughly remains the same for the three quasiparticles. In contrast, the dissipative tensor directly related to the skyrmion Hall angle in Eq. 4 varies a lot due to the distortion of quasiparticles under higher current densities. By increasing the current density from $0.3 \, \text{MA} \cdot \text{cm}^{-2}$ to $3.5 \, \text{MA} \cdot \text{cm}^{-2}$, the value of the $\mathcal{D}_{xx}$ term in the dissipative tensor rises by 0.13%, 14%, and 20.6% for skyrmions, skyrmioniums, and anti-skyrmionites, respectively, which explains well the divergence illustrated in Figs. 2(b) and 2(c). Note that when the applied current densities are large enough ($> 3.5 \, \text{MA} \cdot \text{cm}^{-2}$), we observe the annihilation of skyrmionium and anti-skyrmionite at the edges of the nanotrack. Therefore, we mark the skyrmion Hall angles for the case of the skyrmionium and the anti-skyrmionite with two red cross marks in Fig. 2(c) under such high current densities.

Although multiple quasiparticles have been experimentally demonstrated in liquid crystals [6] and bulk chiral magnets [7], it is vital to explore the conditions in which the skyrmion, skyrmionium, and anti-skyrmionite may stably co-exist in the same system. The co-existence of the three particles proposed in this work is outlined in the stability phase diagram, as shown in Fig. 2(d). The $x$ axis and $y$ axis represent the DMI and PMA constants, respectively. The colour code illustrates the existence/nonexistence of each quasiparticle given a pair of DMI and PMA parameters. There are $31 \times 26 = 806$ data points displayed in Fig. 2(d), and every data point indicates whether any one of three quasiparticles can be stabilised under the corresponding parameters. For each set of the DMI and PMA constants, we configure the system with an initial ansatz that contains a single skyrmion, skyrmionium, and antiskyrmionite, respectively. We let the system equilibrate with the initial states and then mark every data point of Fig. 2(d) with an existence/non-existence for each quasiparticle according to whether the simulation results in the desired equilibrated state.

As shown in Fig. 2(d), the white-coloured region marked with hollow circles represents the stabilisation of the FM state; the yolk-coloured region marked with filled circles is the skyrmion's stabilisation window; the olive region marked with triangles denotes the co-existence of the skyrmion and anti-skyrmionite; the sky-blue area marked with rectangles represents the co-existence of skyrmion and skyrmionium; the navy-coloured region marked with yellow stars is the target parameter window of this work that offers the co-existence of all three quasiparticles. Therefore from Fig. 2(d), it can be summarised that the stabilisation region of the skyrmion is the largest, while the skyrmionium and anti-skyrmionite can stably exist in a subset region of skyrmions. In previous studies reported in the literature e.g. [4,15], only specific type of quasiparticles, such as the skyrmion or the skyrmionium, have been proposed as information carriers in the device. However, in this work, we use multiple skyrmionic quasiparticles simultaneously in a single device, and this proposal is supported by the findings from Fig. 2(d) that the three magnetic quasiparticles can co-exist in a sufficiently wide parameter window for device usage.

## 2.3 Thermal stability of magnetic quasiparticles on the track

To further demonstrate the potential of the proposed VCMA-based interconnect and provide concrete results about the use of multiple quasiparticles simultaneously, we performed a thermal stability analysis with the nudged elastic band method (NEBM) [47], which has been widely used to calculate minimum energy paths (MEPs) of multiple equilibrium states. When performing NEBM, a transition between different local energy minimum states can be visualised as a path with respect to the reaction coordinate, defined by the cumulative sum of the distances between a sequence of configurations along the path. By defining the initial state and the destination state, NEBM will determine the transition path with the minimum energy barrier, i.e., the MEP. It should be noted that NEBM is usually used to calculate MEPs of multiple equilibrium states of the same system [47], and NEBM is especially powerful in searching for MEPs among numerous possible ones. However, in this work, we calculated the energy barrier via NEBM by constraining the target energy path (see Figs. 3(a) and 3(b), also in Fig. 4(a), which will be detailly discussed later) to plausible scenarios. We utilise NEBM here to quantify the probabilities of different cases that elucidate the effect that the synchronisation barriers (notches or VCMA gates) may play, e.g., we compared the energy barriers of the quasiparticles' annihilation in notch-based nanotracks and notch-free nanotracks, which can help decide the safer and stabler synchronisation method for spintronic interconnect. More details of calculating the minimum energy paths (MEPs) and energy barriers can be found in Methods.

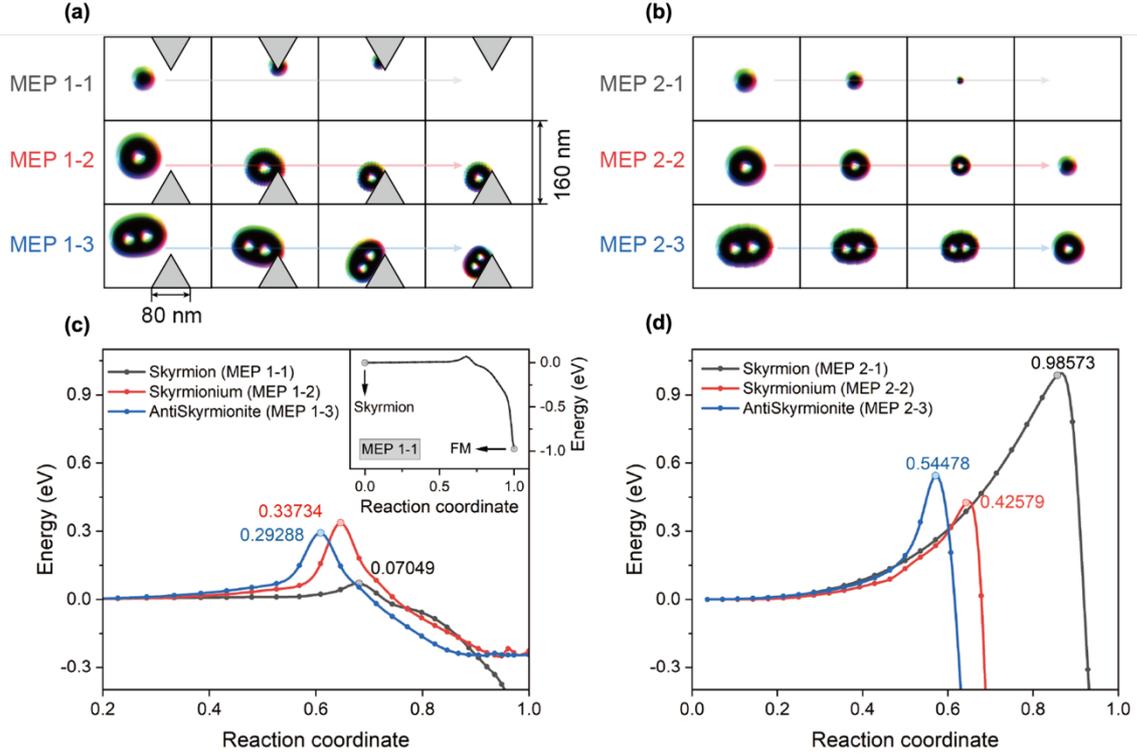

**Figure 3. Minimum energy paths of a skyrmion, skyrmionium, and anti-skyrmionite pinned and annihilation in the nanotrack.** There are two minimum energy paths (MEPs) for the annihilation: (a) a magnetic quasiparticle gets pinned and annihilates at the vertex of the triangle notch, i.e., MEPs 1-1 to 1-3 and (b) a magnetic quasiparticle annihilates in the centre of the nanotrack. Three paths are shown: skyrmion to the ferromagnetic state (MEP 2-1); skyrmionium to skyrmion (MEP 2-2); anti-skyrmionite to skyrmionium (MEP 2-3). (c-d) Energy variation along with the MEPs illustrated in (a-b), respectively. The inset of (c) shows the full MEP 1-1 with the larger range of the $y$-axis illustrating the annihilation of a skyrmion state (energy 0 eV) towards the ferromagnetic state (energy -1 eV). The reaction coordinate is an order parameter representing the relative distance between the states in the configuration space, i.e., 0 stands for the initial state, and 1 represents the destination state. Results displayed in (d) were adapted from Ref. [24]. It should be noted that the energy for the quasiparticles shown in (c-d) is refined with respect to their corresponding initial states in order to facilitate better visualisation and comparison.

We first utilised NEBM in order to estimate the possibility of pinning/annihilation of magnetic quasiparticles when there is a notch in the nanotrack, as proposed in recent numerical studies [10,15,24]. Here, we compared two scenarios, shown in Figs. 3(a) and 3(b), which represent the MEPs of a skyrmion, skyrmionium, and anti-skyrmionite, pinned and annihilated by the triangle notch (MEPs 1-1 to 1-3) and collapsing in the centre of a notch-free nanotrack (MEPs 2-1 to 2-3). We simulated a small section of the nanotrack of 160 nm width, and the notches in the simulations were realised by setting a nonmagnetic equilateral triangle region with a side length of 80 nm, as exhibited in Fig. 3(a). Skyrmionic quasiparticles may also annihilate at the nanotrack edges due to boundary roughness or Magnus force induced by large electrical current densities [13,15]. With regards to edge annihilation of quasiparticles, as often reported in the literature, using high-$\kappa$ materials at the nanotrack edges can also prevent this from happening [10,17,18]. Moreover, the latter concern has already been considered in Sec. 2.2, e.g., by controlling the electrical current density below 4 MA cm$^{-2}$ to prevent the magnetic quasiparticles from annihilating at the edge.

As for devices with a notch-based nanotrack, the annihilation of a magnetic quasiparticle takes place in two stages: it is first pinned by the notch; then it will annihilate on the site. However, our simulation results demonstrate that the skyrmion will immediately annihilate once it gets pinned by the notch, while the skyrmionium and antiskyrmionite show the two-stage annihilation behaviour seen in Fig. 3(a). Although it is also possible for the skyrmionium and antiskyrmionite to be released/depinned from the notch, the pinning itself is sufficient to eliminate the information during the propagation process. Therefore, for the case of the notch-based nanotrack, we consider the energy barrier of magnetic quasiparticles being pinned at the vertex of the triangle notch. This probability can be quantified by determining the energy profile of the MEP illustrated in Fig. 3(a). The calculated energy barrier for the annihilation process of quasiparticles is shown in Fig. 3(c). Among the three quasiparticles, the skyrmion is most likely to annihilate at the notch with the energy barrier of 0.07 eV. Unlike the skyrmion that eventually degrades into the ferromagnetic ground state (see inset of Fig. 3(c)), the skyrmionium and anti-skyrmionite are pinned by the triangle notch afterwards. Although the skyrmionium and anti-skyrmionite exhibit a much more significant energy barrier of 0.34 eV and 0.29 eV, respectively, compared to the skyrmion, it is considerably smaller than the energy barrier for the collapsing process in the notch-free nanotrack shown in Figs. 3(b) and 3(d). This means that under the same amount of current density, magnetic quasiparticles are much more likely to be pinned or annihilate in the notch-based nanotracks than in notch-free ones. Furthermore, we can quantify the stability by analysing the results using the Arrhenius-Néel law to estimate the relaxation time [38,48],

$$\tau(\Delta E) = \tau_0 \exp\left(\frac{\Delta E}{k_B T}\right), \qquad (6)$$

where $f = \tau_0^{-1}$ is attempt frequency, $k_B$ is the Boltzmann constant, $T$ is the temperature under consideration. Here, we assume $T = 300\ K$ as an estimation of quasiparticle lifetime at room temperature. The attempt frequency magnitude is difficult to obtain since the theory typically refers to macrospin systems. In literature, a $10^9$ - $10^{12}$ Hz frequency is typically used [38,48,49]. However, there is a debate about the value of the attempt frequency [50] as it can be as large as $10^{21}$ Hz. Precisely calculating the lifetime of quasiparticles needs a dedicated investigation and is beyond the scope of this paper. We choose a typical value of $10^{12}$ Hz for the attempt frequency here and give an approximate estimation and comparison of the quasiparticle lifetimes. As for MEPs 1-1, 1-2, 1-3, 2-1, 2-2, and 2-3 shown in Fig. 3, we calculate the lifetime of each one using Eq. 6 as 0.015 ns, 458.6 ns, 83.6 ns, $3.66 \times 10^{13}$ ns, $1.43 \times 10^{4}$ ns, and $1.4 \times 10^{6}$ ns, respectively. The results suggest that the magnetic quasiparticles carry a relatively shorter lifetime due to the likely pinning and annihilation at the notches. Therefore, the notch-based nanotrack would be fragile if we employ it in realistic devices, let alone under room temperature thermal fluctuations. In comparison, quasiparticles show a tremendously improved stability in notch-free nanotracks. As shown in Fig. 3(b) and 3(d), quasiparticles exhibit a much longer lifetime in the nanotrack, especially for magnetic skyrmions, whose lifetime skyrockets from 0.015 ns in notch-based nanotrack to $3.66 \times 10^{13}$ ns in notch-free nanotrack. The notches induce an impressive degradation in the device stability by 15 orders of magnitude. Note that the lifetime estimated from energy barriers indicates the thermal stability and annihilation possibility of quasiparticles rather than the precise operation times for realistic devices. In other words, a larger energy barrier contributes to a longer lifetime, which results in better thermal stability of quasiparticles. Therefore, the target of the device design is to obtain larger energy barriers of quasiparticles to withstand greater external disturbances, e.g., thermal fluctuations and electric currents. As a result, quasiparticles will exhibit faster and more reliable operations in the device. In the following, we explore whether VCMA-gated synchronisers in the device provide better thermal stability for magnetic quasiparticles than notches.

As for the notch-free, VCMA-based interconnect device proposed in this work (see Fig. 1), we can roughly estimate the thermal stability of magnetic quasiparticles according to the collapse process illustrated in Fig. 3(b). In this case, the magnetic quasiparticles are most likely to annihilate within VCMA-gated regions or at the corner of the VCMA-gated region and nanotrack edges. The probability of this procedure can be qualitatively described by the stability of quasiparticles against switching into others (i.e., skyrmionium to skyrmion, anti-skyrmionite to skyrmionium, shown in Fig. 3(b)), where all three quasiparticles show larger energy barriers in notch-free devices than notch-based devices. Besides, compared to the strategically etched notches, the VCMA-gated synchronisers proposed in this work have several advantages. First, the VCMA-based interconnect shows better scalability than the notch-based one. Indeed, a single VCMA gate can be used to synchronise multiple quasiparticles, while in the notch-based interconnect device, each magnetic quasiparticle requires one notch artificially etched in a specific position [24]. Second, the VCMA-controlled gate is fully tuneable through voltage and can, therefore, remedy process variations during fabrication. In contrast, notch-based devices are more sensitive to fabrication process variations and, more importantly, are adversely affected by these variations. Therefore, comparison of the energy barrier calculations in Fig. 3(c) and Fig. 3(d) and the reasoning above suggest that magnetic quasiparticles should exhibit better thermal stability and lower probability of annihilation in VCMA-based interconnects than notch-based ones. In the following, we will also investigate the role of VCMA gates.

**2.4 Pipelined spintronic interconnect synchronised by VCMA gates**

By switching VCMA gates on and off in the nanotrack, we effectively manage the lateral energy distribution of the system, which will result in the stop and pass of the information carriers. To obtain the energy distribution of the nanotrack when turning on the VCMA gates, we calculated the energy barrier via NEBM for each quasiparticle when passing a VCMA-gated region where the magnetic anisotropy constant $K_{u1}$ varies. As shown in Fig. 4(a), the width of the nanotrack under simulation is 160 nm and the VCMA-gated region 80 nm. Energy profiles of the skyrmion, skyrmionium, and anti-skyrmionite moving along the track (i.e., MEPs 3-1, 3-2, and 3-3, respectively) were calculated in Fig. 4(b), respectively, where the $K_{u1}$ values of VCMA-gated regions were changed from $0.90K_{u1}$ to $1.10K_{u1}$. Insets of Fig. 4(b) describe the near-linear relation between the energy barrier $E_b$ for quasiparticles crossing the VCMA-gated region and the variation of $K_{u1}$ of the VCMA region. Note that the VCMA range can be as large as $\pm 40\%$ according to [34], but a relatively smaller and feasible range of $\pm 10\%$ is considered in this work.

As for each magnetic quasiparticle, a higher $K_{u1}$ variation ($\Delta K_{u1}$) of the VCMA-gated region leads to more significant energy differences in and out of the VCMA-gated region. Comparing the same amount of $\Delta K_{u1}$ with a positive and negative sign, the negative one leads to a higher energy barrier. The difference in energy barriers of $\pm \Delta K_{u1}$ could be attributed to the shape shrinking/expanding of quasiparticles when the PMA constant is adjusted, in addition to the energy change purely due to the PMA. As a result, the calculated energy barrier for the $+\Delta K_{u1}$ VCMA gate is reduced, while the energy barrier for the $-\Delta K_{u1}$ VCMA gate is strengthened. Such asymmetry of energy barriers may result in similar asymmetric effects on device functionality where the VCMA gates are deployed. These results are in good agreement with other simulations on the pinning/depinning properties of individual skyrmions via VCMA gates [14,51]. Therefore, the VCMA-gated regions with lower $K_{u1}$ could potentially serve as registers to store state information of quasiparticles. Among these three quasiparticles, the skyrmion experiences the lowest energy barrier passing the VCMA gate than the skyrmionium and the anti-skyrmionite under the same $\Delta K_{u1}$. For example, if we change PMA to be positive 10% by applying a voltage on the VCMA gate, the energy barrier for skyrmion, skyrmionium and anti-skyrmionite crossing the region is calculated as 0.325 eV, 0.813 eV, and 1.158 eV, respectively. However, this does not mean that skyrmions are more likely to cross over the VCMA-gated region than skyrmioniums and anti-skyrmionites under the same amplitude of current density, because the additional energy due to the current-induced spin-transfer torques is different for each quasiparticle. In fact, from the micromagnetic simulations, we noted that by increasing the current density in the nanotrack, the anti-skyrmionite is the first to cross the VCMA-controlled barrier while the other two are stuck.

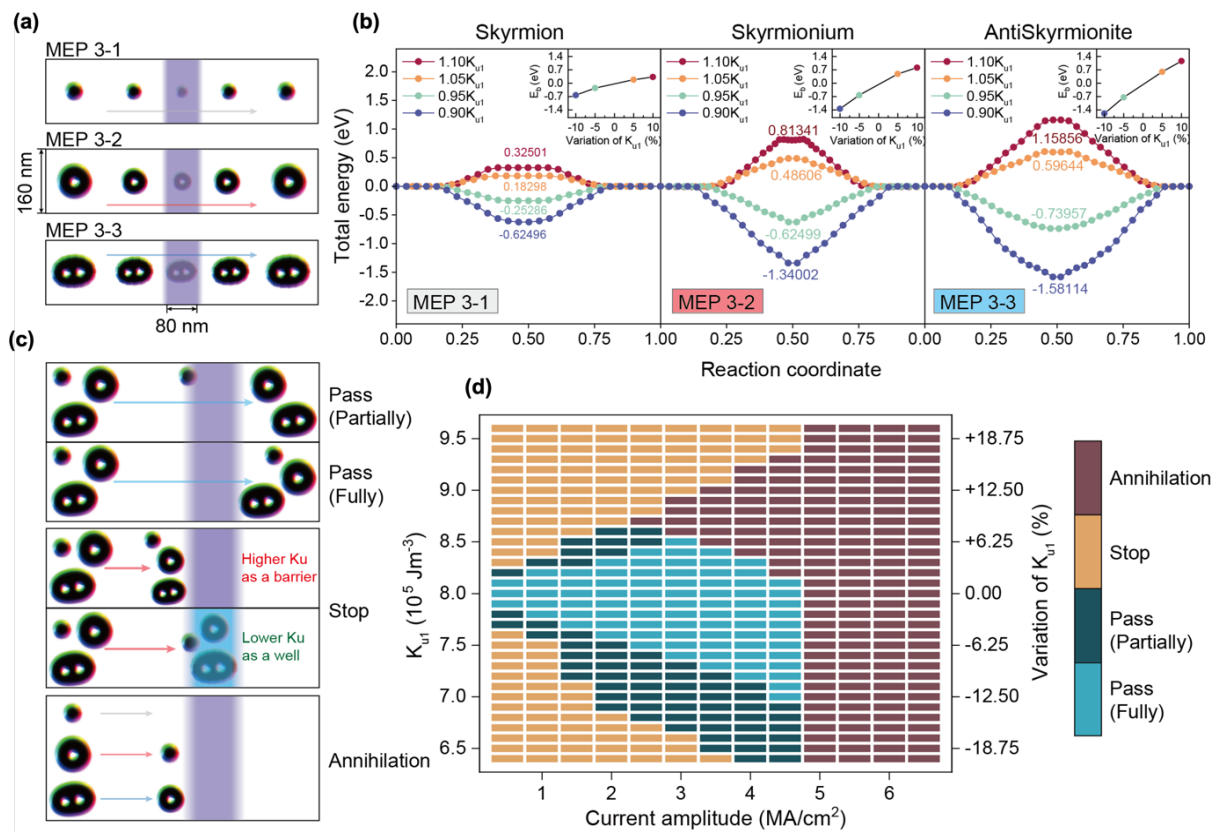

**Figure 4. Phase diagram of the VCMA-gated region for skyrmion, skyrmionium, and anti-skyrmionite.** (a) Illustrations of MEPs of a skyrmion, skyrmionium, and anti-skyrmionite passing through a VCMA region. (b) MEPs of the three magnetic quasiparticles are shown in (a) calculated using the NEBM. Insets illustrate the energy barrier $E_b$ with varying $K_{u1}$ of the VCMA-gated region. (c) Schematic illustration of the skyrmionic quasiparticles passing the VCMA-gated region in the nanotrack where the purple shade represents the VCMA barrier, and the sky-blue shade stands for VCMA well. The quasiparticles propagate along the nanotrack when the constant driving current is applied and the VCMA gate is turned on. There exist four cases: i) pass partially, ii) pass entirely, iii) stop, and iv) annihilation of the quasiparticles. (d) The working window of the skyrmion, skyrmionium, and anti-skyrmionite with different current densities and magnetic anisotropy constant $K_{u1}$ was calculated with systematic micromagnetic simulations.

As discussed above, we can achieve either a higher or lower PMA and energy barrier by applying a positive or negative voltage to the VCMA gates. As shown in Fig. 4(c), the region with higher PMA serves as an energy barrier to stop the magnetic textures (for some transition values, one or two of the particles would pass the VCMA-gated region, which we refer to as a partial pass here) and the region with lower PMA offers a potential well to trap the quasiparticles. From the results of Fig. 4(b) we observe, if the applied current density is smaller than the critical value that compensates the energy barrier/potential well, quasiparticles would be stopped by the VCMA gate. Otherwise, the quasiparticles will pass the VCMA-gated region (see top panel in Fig. 4(c)). Furthermore, if the energy of the applied current is larger than the energy barriers shown in Figs. 3(b) and 3(d), the quasiparticles will annihilate during propagation.

There are four possible cases: i) pass partially, ii) pass fully, iii) stop, and iv) annihilation, depicted in Fig. 4(c). We then performed a series of simulations of a nanotrack with a VCMA gate placed in the middle of the track. By scanning the $K_{u1}$ value of the VCMA-gated region and the current amplitude injected in the nanotrack, we obtained the working window for the pass/stop/annihilation status of quasiparticles in the device, as shown in Fig. 4(d). We considered the simulation containing all three quasiparticles rather than having each one individually. Because in real devices, we use three of them together, we need to guide the device design with the strictest condition where all three particles need to cross or be stopped by the VCMA gate. The results organised in Fig. 4(d) fit well with our prediction in the discussion above. Under smaller current densities, the VCMA gate with high $\Delta K_{u1}$ will stop the quasiparticles, while quasiparticles will pass the VCMA gate with lower $\Delta K_{u1}$. Above high current densities ($> 4.5 \text{ MA} \cdot \text{cm}^{-2}$), the annihilation of quasiparticles will be seen. Another significant result in Fig. 4(d) is the asymmetry of the "annihilation" phase along with $K_{u1}$, which could be explained by the asymmetry of the energy barrier $E_b$ of each quasiparticle with $K_{u1}$ in Fig. 4(b). For negative $\Delta K_{u1}$, the quasiparticles find it difficult to escape from the VCMA well due to the higher $E_b$. So, in the bottom half of Fig. 4(d), the quasiparticles will not annihilate until the current density increases over $4.5 \text{ MA} \cdot \text{cm}^{-2}$, which is the threshold value to prevent quasiparticles from collapsing.

Based on the results discussed above, we carefully designed a scheme for a pipelined spintronic interconnect, as shown in Fig. 5. The length of the nanotrack is 4800 nm, and the width is 240 nm. As schematically illustrated in Fig. 1, the device has a three-branch encoder and a three-branch decoder to encode and decode three sequences of information signals. In the simulation, we placed eight VCMA gates on the main track and several VCMA gates near the nucleation and detection heads in the branches. As for the setup of the VCMA effect, we set a 5% variance of $K_{u1}$ for VCMA gates during the simulation. It should be noted that the number of VCMA gates employed in the device is determined by the number of bits. Here, we want to examine the transmission of one byte (eight bits) information, which is usually the smallest addressable unit of memory in many computer architectures. Moreover, due to the non-volatility of skyrmionic quasiparticles, such a design can also be used to implement an 8-bit shift register.

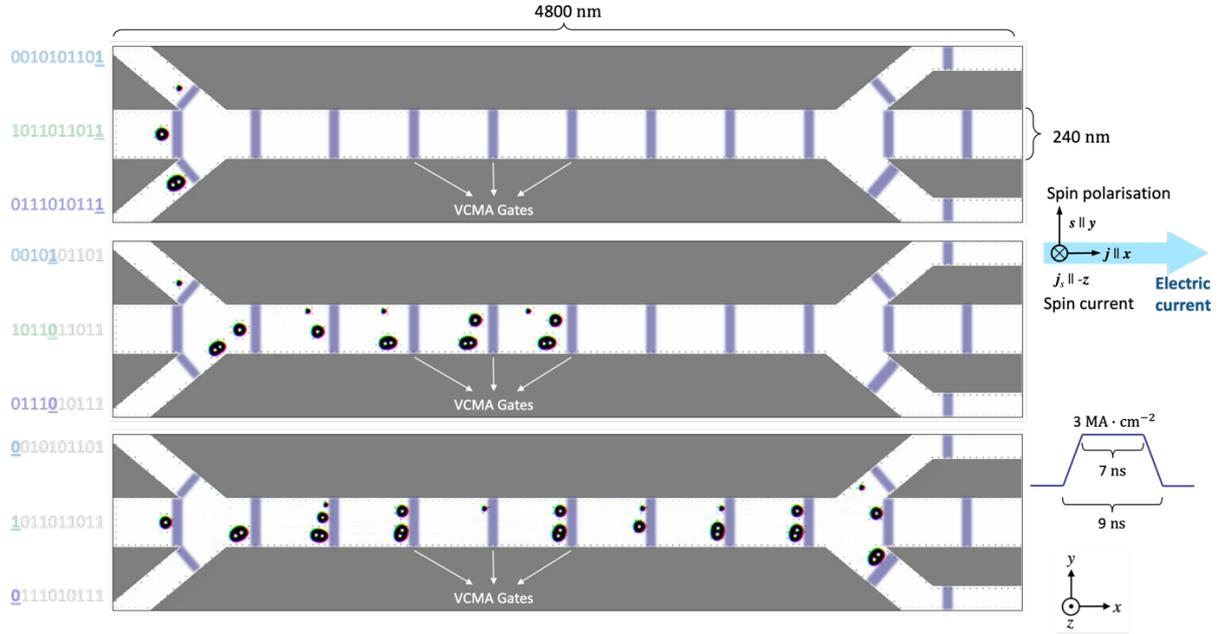

**Figure 5. Micromagnetic simulations of the pipelined spintronic interconnect.** The racetrack is divided into separate regions via VCMA gates. The device is powered by a series of electrical pulses perpendicular to the plane (CPP) with an amplitude of $3$ MA cm$^{-2}$ for $9$ ns. The pulse consists of 1 ns rising edge, 7 ns constant current, and 1 ns falling edge. The width of the main racetrack is $240$ nm and the length of the device $4800$ nm, and the spacing between adjacent VCMA gates is $400$ nm.

The presence of the magnetic quasiparticles encodes logic "1", while the absence of the quasiparticles corresponds to logic "0". A series of driving current pulses is periodically applied in the HM layer in the direction of $x$-axis with 1 ns rising edge and 1 ns falling edge. The spin current is therefore injected perpendicular to the FM plane (in the direction of $-z$) with the spin polarisation in $+y$ direction. We include rising and falling edges in our pulses because we want to provide realistic operating conditions of current injection for experiments/devices. At the same time, the rising edge of current pulses can provide the device with an initialising process, and the falling edge can ensure all the quasiparticles reach and are stopped by the VCMA-gated synchroniser simultaneously to protect the order of information sequences. In the simulation shown in Fig. 5, we send and fully transmit three sequences of information signals "1011010100 (blue), 1101101101 (green), 1110101110 (purple)" serially. In Fig. 5, the information sequences are printed in reverse order. Information bits are multiplexed, transmitted through the pipeline, and demultiplexed simultaneously, tripling the bandwidth of an interconnect that carries only a single information carrier (i.e., a magnetic particle). Therefore, with the help of VCMA gates, the proposed skyrmionic interconnect inherently supports data pipelining, where the interconnect throughput is boosted, despite of a 10-pulse long latency for the first piece of data to be received. Note that such latency is given in the context of the maximum interconnect throughput, where the electric pulses are sent after previous one immediately without rest time. These results illustrate potential applications of our proposed spintronic interconnect device.

## 3. Discussion

### 3.1 Tuneable interconnect performance and analysis on the energy-efficiency

We have demonstrated the whole device operation flow, and the pipelined scheme of the voltage-controlled spintronic interconnect device. To evaluate the possible benefits of our proposed interconnect for future integrated systems, we describe the tunability of the device performance and compare its energy efficiency with copper interconnects, which are commonly used in the mainstream CMOS technology.

The performance of interconnects can be quantified by their maximum throughput as well as their energy efficiency. The throughput of the pipelined interconnect introduced in Fig. 5 is given by [52],

$$X = \frac{C}{\tau}, \tag{7}$$

where $X$ is the maximum throughput of the proposed interconnect device and $C$ is the total amount of transferred data bits within the time $\tau$. As indicated from Fig. 6(a), the device throughput can be effectively tuned by adjusting current densities. Because the skyrmionium and antiskyrmionite annihilate at $J > 4 \text{ MA} \cdot \text{cm}^{-2}$, the current densities here are limited to $4 \text{ MA} \cdot \text{cm}^{-2}$, where $\tau = 7.5$ ns is obtained from the simulations to propagate an information package between adjacent VCMA gates. This sets an upper limit of 400 Mbps maximum throughput of the device (calculated via Eq. 7 with $C = 3$ bits), shown in Fig. 6(a). Regarding the interconnect latency, there is a 10-pulse long latency between the two ends (i.e., sender and the receiver), i.e., 90 ns calculated by the electrical pulse utilised in the pipelined scheme of Fig. 5. At the same time, this latency can also be tuned by changing the current densities. For example, if we want to achieve a 10-time higher interconnect throughput we increase the current density of the applied pulses such that the quasiparticles propagate faster, and the required pulse width is shortened by 10 times. Therefore, the latency will be correspondingly reduced by 10 times. As shown in Fig. 6(a), there is an almost linear relationship between the maximum device throughput and current densities, which indicates the effective tunability of the device performance merely by modifying the current supply.

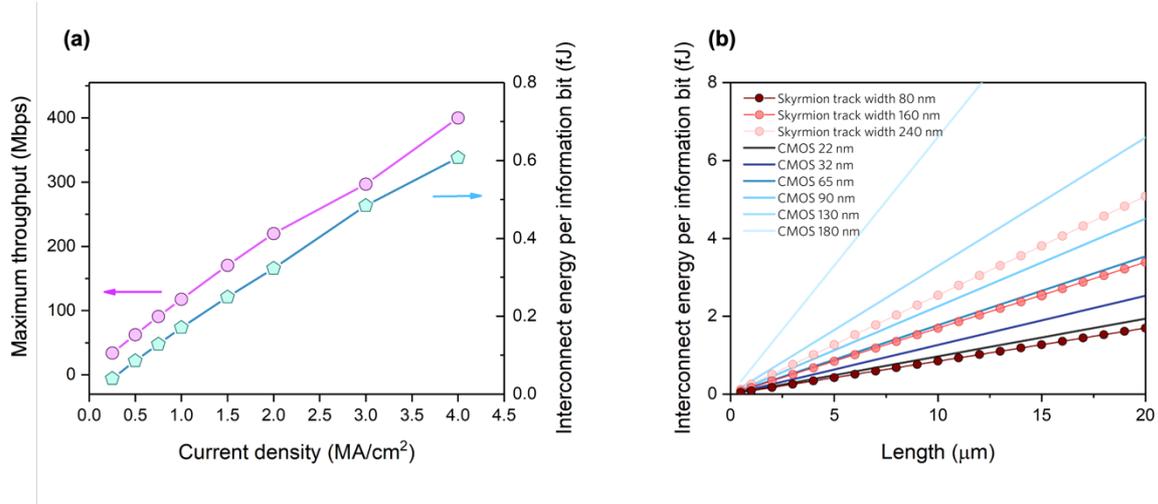

**Figure 6. Tunability of the device performance and energy consumption.** (a) Maximum throughput and energy consumed for single information bit transmission of the proposed pipelined device under different current densities. (b) Comparison of the energy efficiency of the proposed pipeline spintronic interconnects with the copper interconnects in CMOS technology nodes varying from 180 nm to 22 nm fabrication process.

The energy consumption for information transmission of the proposed spintronic interconnect is given by

$$E = \rho t l w J^2 T, \tag{8}$$

where $E$ is the energy consumption, $\rho$ is the resistivity of the HM layer, $t$ is the thickness of the HM layer, $l$ is the length of the racetrack, $w$ is the width of the nanotrack, $J$ is the density of the charge current applied in the HM layer, and $T$ is the pulse duration for the proposed pipelined spintronic interconnect. The values of the parameters are chosen as follows: $l = 400$ nm (distance between two VCMA gates); $t = 3.8$ nm (the spin diffusion length of Pt [40,53]); $w = 240$ nm; $\rho = 50$ μΩcm for 3.8 nm Pt thin film [40]. We integrate $J^2 T$ with the current pulse shape depicted in Fig. 5. The transmission energy per information bit with respect to current density is also shown in Fig. 6(a). Similar to the linear tunability of the interconnect maximum throughput, the transmission energy rises linearly with respect to the applied current as well. Results in Fig. 6(a) provide design guidelines for future implementations of such interconnect devices, which can tune the device performance within the power limitations in practical devices and application scenarios.

Finally, we would like to evaluate the spintronic interconnect by comparing it with conventional CMOS technology to offer better insight into the suitability of the skyrmionic interconnect devices. We calculate the energy consumption of the copper interconnects across different CMOS technology nodes. In CMOS interconnects, the energy consumption to transfer a bit via a copper interconnect is described by

$$E = C_{\text{wpu}} l V_{\text{dd}}^2, \qquad (9)$$

where $C_{\text{wpu}}$ is the wire capacitance per unit length, $l$ is the length of the copper wire and $V_{\text{dd}}$ is the voltage supply. For a fair comparison, we choose length $l = 400$ nm, the same with the spintronic interconnect. Copper wire capacitance $C_{\text{wpu}}$ and the voltage supply $V_{\text{dd}}$ of different CMOS technologies are estimated via the Predictive Technology Model (PTM) [54]. The detailed parameters of copper interconnect and calculation via PTM can be found in Methods. Fig. 6(b) compares the transmission energy with respect to interconnect length for the proposed spintronic interconnect device and the copper interconnect in 180 nm down to 22 nm CMOS technology nodes. As for the spintronic interconnect, we choose the transmission energy for a 100 Mbps throughput from Fig. 6(a). As shown in Fig. 6(b), the estimated energy efficiency of the spintronic interconnect with a 240 nm width nanotrack is comparable with (slightly better than) a copper interconnect in the 90 nm CMOS technology node. By narrowing the width of the skyrmion nanotrack, transmission energy can be further reduced. We have also calculated the transmission energy of the spintronic interconnect devices whose widths are 160 nm and 80 nm, respectively, by assuming that the antiskyrmionite would be stabilised at the 80 nm width nanotrack. The 160 nm width spintronic interconnect exhibits a similar energy efficiency with copper interconnect in 65 nm CMOS node, while the 80 nm width spintronic interconnect presents comparable energy efficiency with that of the 22 nm CMOS node. According to Eq. 8, reducing the width of the nanotrack $w$ can directly decrease the data transmission energy since the calculated energy $E$ is linearly dependent on $w$. However, the width of the nanotrack cannot be narrower than the lateral dimensions of magnetic quasiparticles, which varies in different material systems. In this work, the diameters of the skyrmion, skyrmionium, and anti-skyrmionite (minor diameter of its ellipse shape) are estimated to be approximately 30 nm, 60 nm, and 80 nm. Consequently, we set 80 nm as the lower limit of the nanotrack width $w$.

The information transmission energy of the spintronic interconnect device determined here also includes the multiplexing and demultiplexing procedures, while the calculated energy for copper interconnects purely accounts for bit transmission without considering any control signal. At the same time in spintronic interconnects additional energy is spent for switching on/off the VCMA gates that guide the quasiparticles along the nanotrack, which can be implemented by periphery circuits. For simplicity, we assume the energy to switch VCMA gates in spintronic interconnect is comparable to the energy to send control signals in CMOS interconnects. Therefore, we use the Eqs. 8 and 9 to directly compare data propagation energy between spintronic and copper interconnect in CMOS, as shown in Fig.

6(b). The concept of spintronic interconnects with multiple quasiparticles is even more advantageous since data transmission is accompanied by automatic multiplexing and demultiplexing at both ends. The non-volatility and bidirectionality of the devices are also noteworthy benefits. We anticipate that future work on spintronic interconnect devices could include i) evaluating spintronic interconnect devices in realistic conditions, e.g., magnetic multilayer structures with realistic material grains and defects at room temperature, ii) research on how to enhance the thermal stability of quasiparticles in the device, and iii) investigating strategies to expand throughput for given energy efficiency and iv) experimental realisations.

## 4. Conclusions

In this work, we propose to utilise multiple magnetic quasiparticles as information carriers (i.e., magnetic skyrmion, skyrmionium, and anti-skyrmionite) in a spintronic interconnect device based on voltage-controlled magnetic anisotropy (VCMA) gates. The device can achieve automatic multiplexing/demultiplexing and simultaneous transmission of multiple information signals. Combining theoretical analysis with micromagnetic simulations, we demonstrated the distinct current-driven behaviour of different magnetic quasiparticles. Through the NEBM, we showed that quasiparticles in VCMA-based interconnect are more thermodynamically stable than notch-based structures. A pipelined interconnect is then illustrated by embedding VCMA-based gates as synchronisers. Lastly, we show that the energy efficiency of the skyrmionic interconnect is comparable to copper interconnects in CMOS technologies. This work significantly widens the possibilities for all-magnetic spintronic devices with multiple quasiparticles as information carriers and should be relevant in a future potential holistic spintronic nanocomputing paradigm.

## Methods

**Micromagnetic simulation**

The micromagnetic simulations were performed using the GPU-accelerated micromagnetic programme MuMax[3] [37]. The time-dependent magnetisation dynamics are conducted by the Landau-Lifshitz-Gilbert (LLG) equation with an additional term accounting for the Slonczewski spin-orbit torque [55]:,

$$\frac{d\mathbf{m}}{dt} = -\gamma_e \mathbf{m} \times \mathbf{B}_{\text{eff}} + \alpha \mathbf{m} \times \frac{d\mathbf{m}}{dt} + \mathcal{T}_{\text{SOT}}[\mathbf{m} \times (\mathbf{m}_p \times \mathbf{m})], \quad (10)$$

where $\mathbf{m} = \mathbf{M}/M_s$ is the reduced magnetisation, $M_s$ is the saturation magnetisation, $\gamma_e = 1.76 \times 10^{11} \text{ T}^{-1}\text{s}^{-1}$ is the gyromagnetic ratio of an electron, $\mathbf{B}_{\text{eff}}$ is the space- and time-dependent effective magnetic field, $\alpha$ is the dimensionless Gilbert damping parameter, $\mathcal{T}_{\text{SOT}} = \frac{\gamma_e \hbar J_e \theta_{\text{SH}}}{2eM_s t}$ is the SOT efficiency, where $\hbar$ is the reduced Planck constant, $J_e$ is the electron current density, $\theta_{SH}$ is the spin Hall angle, $e$ is the electron charge, $t$ is the thickness of FM layer, and $\mathbf{m}_p$ is the polarisation direction of the spin current due to the spin Hall effect. The micromagnetic energy density $\varepsilon(\mathbf{m})$ is a function of $\mathbf{m}$, which contains the Heisenberg exchange energy term, the anisotropy energy term, the Zeeman energy term, the magnetostatic energy term and the DMI energy term. The material parameters to perform the simulations are chosen following Refs. [4,11,15]: damping parameter $\alpha = 0.3$, interfacial DMI constant $D_{int} = 3.5$ mJ·m$^{-2}$, saturation magnetization $M_s = 580$ kA·m$^{-1}$, the spin Hall polarisation $\Theta_{SH} = 0.6$ to enhance the spin Hall effect, the uniaxial out-of-plane magnetic anisotropy $K_{u1} = 800$ kJ·m$^{-3}$, the polarisation of the spin current is in the $+y$ direction, and the exchange constant is assumed to be $A = 15$ pJ·m$^{-1}$. To ensure the accuracy of calculation, the mesh size of discretisation is set to $1 \times 1 \times 1$ nm$^3$, which is much smaller than the exchange length $l_{ex} = \sqrt{2A/(\mu_0 M_s^2)} = 8.42$ nm and DMI length $l_{DMI} = 2A/D_{int} = 8.57$ nm. The thickness of the heavy metal layer is 3.8 nm which is chosen as the spin diffusion length of Pt [40]. An external magnetic field of 10 mT in the out-of-plane direction is applied. The edges of 5 nm thickness with higher magnetic anisotropy $K_{u1,high} = 900$ kJ·m$^{-3}$ is set for the device to avoid magnetic quasiparticles annihilate at the nanotrack edges.

**The solution to the Thiele equation in the presence of the SOT**

As introduced in the main text, there are two situations to be considered: 1) propagation of the magnetic quasiparticle far from the edge; and 2) the quasiparticle eventually moving along the direction of the applied current along the nanotrack edge. By calculating the dissipative tensor $\mathcal{D}$ and the driving torque tensor $\mathcal{J}$ via MATLAB using the micromagnetic profiles of the skyrmion, skyrmionium, and anti-skyrmionite, the tensor $\mathcal{D}$ and $\mathcal{J}$ have the following shapes,

$$\mathcal{D}_{xx} = \mathcal{D}_{yy} = \mathcal{D}, \quad \mathcal{D}_{xy} = \mathcal{D}_{yx} \approx 0, \tag{11}$$

$$\mathcal{J}_{xy} = -\mathcal{J}_{yx} = \mathcal{J}, \quad \mathcal{J}_{xx} = \mathcal{J}_{yy} \approx 0. \tag{12}$$

Here we assume that the magnetic quasiparticle does not perform displacement in $z$ direction. For a magnetic quasiparticle propagating in a nanotrack with periodical boundary conditions along $x$ ($U(\mathbf{r}) = U(y)$), the Thiele equation in Eq. 3 reads,

$$-4\pi \mathcal{N}_{sk} \begin{pmatrix} -v_y \\ v_x \end{pmatrix} = \alpha \mathcal{D}_{xx} \begin{pmatrix} v_x \\ v_y \end{pmatrix} + \mathcal{T}_{SOT} \mathcal{J}_{xy} \begin{pmatrix} m_{py} \\ -m_{px} \end{pmatrix} + \begin{pmatrix} 0 \\ \partial_y U(y) \end{pmatrix}. \tag{13}$$

For the magnetic quasiparticles far from the edge, $\partial_y U(y) = 0$. Assuming the injected spins along $-y$ direction, i.e., $m_{px} = 0$, $m_{py} = -1$, Eq. 13 can be further simplified to,

$$-4\pi \mathcal{N}_{sk} \begin{pmatrix} -v_y \\ v_x \end{pmatrix} = \alpha \mathcal{D}_{xx} \begin{pmatrix} v_x \\ v_y \end{pmatrix} - \mathcal{T}_{SOT} \mathcal{J}_{xy} \begin{pmatrix} 1 \\ 0 \end{pmatrix}. \tag{14}$$

Solving Eq. 14, we can extract the Hall angle of the particle movement as,

$$\theta_{\text{SkHE}} = arctan(\frac{v_y}{v_x}) = -arctan(\frac{4\pi \mathcal{N}_{\text{sk}}}{\alpha \mathcal{D}_{\text{xx}}}). \qquad (15)$$

For the magnetic quasiparticles steadily moving along the track edge ($v_y = 0$), Eq. 13 can be written as,

$$-4\pi \mathcal{N}_{\text{sk}} \begin{pmatrix} 0 \\ v_x \end{pmatrix} = \alpha \mathcal{D}_{\text{xx}} \begin{pmatrix} v_x \\ 0 \end{pmatrix} - \mathcal{T}_{\text{SOT}} \mathcal{J}_{\text{xy}} \begin{pmatrix} 1 \\ 0 \end{pmatrix} + \begin{pmatrix} 0 \\ \partial_y U(y) \end{pmatrix}. \qquad (16)$$

We obtain the velocity of the magnetic quasiparticle in $x$ direction,

$$v_x = \frac{\mathcal{T}_{\text{SOT}} \mathcal{J}_{\text{xy}}}{\alpha \mathcal{D}_{\text{xx}}} = \frac{\gamma_e \hbar \mathcal{J}_{\text{xy}} J_e \theta_{SH}}{2e M_s t \alpha \mathcal{D}_{\text{xx}}}. \qquad (17)$$

The results from the micromagnetic simulations and the Thiele equation are summarised in Table 1 below.

**Table 1.** Results of the simulations and theoretical predictions with Thiele equation.

| Magnetic quasiparticle | $\mathcal{N}_{\text{sk}}$ | $\mathcal{N}_{\text{sk}}$ MATLAB | $\mathcal{D}_{\text{xx}}$ Thiele | $\mathcal{J}_{\text{xy}}$ Thiele (nm) | $\theta_{\text{SkHE}}$ Sim. (Deg.) | $\theta_{\text{SkHE}}$ Thiele (Deg.) | $v_x$ Sim. (m/s) |
|---|---|---|---|---|---|---|---|
| Skyrmion | -1 | -0.9998 | 22.58 | 148.8 | 61.5 | 61.66 | 49.7 |
| Skyrmionium | 0 | -0.0002 | 64.89 | 429.8 | -0.2 | 0 | 55.0 |
| Antiskyrmionite | 1 | 0.9995 | 102.76 | 676.26 | -26.3 | -24.88 | 51.4 |

**Predictive technology model**

The Predictive Technology Model (PTM) can provide accurate, customisable, and predictive model files for transistor and interconnect technologies [54]. It is compatible with standard circuit simulators (e.g., SPICE) and scalable with disparate process variations. PTM is broadly used for pathfinding activities before a semiconductor technology is fully developed. Therefore, it is an ideal tool to help us calculate the information transmission energy for CMOS technologies. To calculate the transmission energy via Eq. 9, we need the value of parameters $C_{\text{wpu}}$ and $V_{\text{dd}}$. The voltage supply $V_{\text{dd}}$ is looked up from Ref. [56], and the dimensions are estimated from the technology sheets in Ref. [57]. The wire capacitance per unit length of the copper interconnect $C_{\text{wpu}}$ can be estimated by using the PTM with salient parameters of the copper interconnect across several generations of CMOS technology including, for instance, width, space, thickness, height, and dielectric. The typical values of the above parameters for different generations of CMOS technology are summarised in Table 2. The calculated $C_{\text{wpu}}$ and transmission energy per information bit are also listed in Table 2.

**Table 2.** Parameters of the copper interconnect in various generations of CMOS technology utilised in the PTM and the calculated wire capacitance and transmission energy per information bit.

| CMOS technologies | $V_{dd}$ (V) | Width (nm) | Space (nm) | Thickness (nm) | Height (nm) | Dielectric $\kappa$ | $C_{wpu}$ (fF/mm) | Energy (fJ) Length 400 nm |
|---|---|---|---|---|---|---|---|---|
| 22 nm | 0.9 | 35 | 35 | 70 | 70 | 1.8 | 119.62 | 0.04845 |
| 32 nm | 1.0 | 50 | 50 | 100 | 100 | 1.9 | 126.22 | 0.06311 |
| 65 nm | 1.1 | 100 | 100 | 200 | 200 | 2.2 | 146.20 | 0.08845 |
| 90 nm | 1.1 | 150 | 150 | 300 | 300 | 2.8 | 186.08 | 0.11258 |
| 130 nm | 1.2 | 200 | 200 | 450 | 450 | 3.2 | 228.66 | 0.16464 |
| 180 nm | 1.8 | 280 | 280 | 650 | 650 | 3.5 | 255.32 | 0.33009 |

**Minimum energy path calculations**

Nudged elastic band method (NEBM) [47] has been widely used to calculate minimum energy paths (MEPs) of multiple equilibrium states (energy minima). A transition between different states can be visualised as a path with respect to the reaction coordinate, defined by the cumulative sum of the distances between a sequence of configurations along the path. The path that requires the minimal energy cost is referred to as a minimum energy path, and an energy barrier $E_b$ of the transition is calculated by the difference between an energy maximum (saddle point) and an energy minimum. Different from the Monte Carlo method, which samples the most probable transition paths, the NEBM starts from an initial guess of a path, and the algorithm minimises the path by lowering the saddle point, by analogy with tensioning an elastic band across a mountain. In this work, micromagnetic simulations of this part are performed using Fidimag [38] for calculations of equilibrium states of topological spin textures, where magnetisations are relaxed based on the Landau-Lifshitz equation of motion (Eq. 10), followed by the minimisation of the total energy by the steepest descent method [58]. Magnetic parameters utilised in Fidimag is the same as those in mumax3. The NEBM method is then used to calculate MEPs between the proposed equilibrium states to quantify the energy barriers along with the transitions.

## Acknowledgements

R.C. and Y.L. wish to acknowledge the China Scholarship Council (CSC) and the Department of Computer Science Kilburn Scholarship for the funding support. R.C. would like to thank the University of Manchester President's Scholarship. The authors would also like to acknowledge the assistance provided by Research IT and the use of the Computational Shared Facility at the University of Manchester. This work is supported by the Engineering and Physical Sciences Research Council (EPSRC) under the grant 'Skyrmionics for Neuromorphic Technologies', EP/V028189/1.

## Data availability

The data that support the plots within this paper and other findings of this study are available from the corresponding authors upon reasonable request.

## Author contributions

R.C., C.M., and V.P. conceived the project. R.C. performed the micromagnetic simulations and energy calculations. Y.L. performed the nudged elastic band method (NEBM) for energy barriers. R.C. and Y.L. prepared the manuscript. All authors discussed and commented on the analysis and the manuscript. All authors have approved the final version of the manuscript.

## Competing interests

The authors declare no competing interests.

## Additional information

**Reprints and permissions information** is available at [www.nature.com/reprints](www.nature.com/reprints).
**Correspondence and requests for materials** should be addressed to R.C. or C.M.
**Publisher's note**: Springer Nature remains neutral with regard to jurisdictional claims in published maps and institutional affiliations.